\documentclass[12pt]{aa}
\usepackage[english]{babel}
\usepackage{psfig,graphicx}
\usepackage{longtable}

\onecolumn
\newcommand{\ab}{Astrophys. Bull. }
\newcommand{\arep}{Astron. Rep. }
\newcommand{\alet}{Astron. Let. }
\newcommand{\araa}{Ann. Rev. Astron. Astrophys. }
\newcommand{\mnras}{Mon. Not. R. Astron. Soc. }
\newcommand{\apj}{Astrophys. J. }
\newcommand{\apjs}{Astrophys. J. Suppl.}
\newcommand{\aj}{Astron. J. }
\newcommand{\aaa}{Astron and Astrophys.}
\newcommand{\aas}{Astron and Astrophys. Suppl.}
\newcommand{\pasp}{Publ. Astron. Soc. Pasif. }

\begin{document}

\title{Circumstellar envelope manifestations in the optical spectra  of evolved stars}

\author{V.G. Klochkova}

\institute{Special Astrophysical Observatory RAS, Nizhnij Arkhyz,  369167 Russia}

\date{\today} 

\abstract{We consider the peculiarities of the optical spectra of far evolved stars with circumstellar 
 gaseous--dusty envelopes: the time variability of the absorption-emission profiles of the H$\alpha$ line,
the presence of stationary emission and absorption molecular bands, multicomponent complex 
profiles of the Na\,I D--doublet lines. We show that the peculiarities
of the line profiles (the presence of an emission component in the Na\,I D doublet lines,
the specific type of the molecular features, the asymmetry or splitting of the profiles of strongest
absorptions with low excitation potential of the low level) can be associated with the
kinematic and chemical properties of the circumstellar envelope and its morphological type.}

\titlerunning{\it Circumstellar envelope  in the optical spectra of evolved stars }
\authorrunning{\it Klochkova}

\maketitle

\section{INTRODUCTION}

In this paper we analyze the features of the optical spectra of far evolved stars that form in 
extended gas and dust envelopes. The main targets of the studied sample are  protoplanetary nebulae
(PPN)---objects at the post-asymptotic branch (post-AGB) stage of evolution, as well as some related 
stars with large infrared excesses. On the Hertzsprung--Russell diagram stars undergoing the
short-lived PPN stage evolve from the asymptotic giant branch toward the planetary nebula (PN) 
stage at almost constant luminosity, getting increasingly hotter in the process. These
descendants of AGB stars are low-mass cores with typical masses of $0.6\,\mathcal{M}_{\odot}$ 
surrounded by an extended and often structured gaseous envelope, which formed as a result of
substantial mass loss by the star during the preceding evolutionary stages. AGB and post-AGB stars 
are popular among both theorists and observers because it  is during these evolutionary
stages that the synthesis and subsequent dredge-up (third mixing) of carbon and heavy s--process 
metals~[\cite{Busso,Herwig}] occur. AGB stars are therefore the principal suppliers of heavy metals and
important suppliers of carbon and nitrogen to the interstellar medium~[\cite{Gustaf}],  
thereby participating in the chemical evolution of galaxies.

Furthermore, AGB and post-AGB stars also owe their popularity to the fact that they can serve as tools 
for the study of stellar-wind manifestations. The mass loss of evolved stars is the dominating factor 
of the final stages of  stellar evolution. The specifics of mass loss  (mass-loss rate, sequence of wind
episodes, interaction of stellar winds, and details of the chemical composition of the upper outflowing layers) 
determine the chemical composition and structure of the circumstellar envelopes that form around protoplanetary 
and planetary nebulae. The stellar-wind history thus remains recorded in the shape,
structural features, and chemical peculiarities of the circumstellar envelope.

However, the details of the mass loss process during the AGB--PN transition remain unclear. This is primarily 
true for the physical processes that shape the complex structure of PPN envelopes. Images taken with the Hubble Space
Telescope~[\cite{Ueta2000,Sahai}] and during space IR missions~[\cite{Cox}] are rarely spherical.
Central stars are usually surrounded by envelopes in the form of extended haloes,  arcs, lobes, and tori. Some stars exhibit
various combinations of the above features as well as bipolar and quadrupolar nebulae with dust bars. Examples of the latter
two types include the Egg\,=\,RAFGL\,2688 and IRAS\,19475+3119 nebulae, for which high-resolution images
were taken by the Hubble Space Telescope~[\cite{Siodmiak}]. All PPNe and about  80\% of
planetary nebulae are asymmetric~[\cite{Lagadec}]. Much remains to be understood about the processes of condensation of
dust particles and formation of the dust fraction in AGB star envelopes (see~[\cite{Bieging}] and references therein).

\renewcommand{\baselinestretch}{1.0}

A circumstellar gas and dust envelope shows up in the form of peculiarities in the IR, radio, and optical spectra of post-AGB
supergiants. The optical spectra of PPNe differ from those of classical massive supergiants by the presence of molecular bands
superimposed onto the spectrum of an F--G supergiant and by the anomalous behavior of the profiles of selected spectral features.
These may include complex emission and absorption profiles of H\,I, Na\,I, and He\,I lines, profiles of strong absorptions
distorted by emissions or splitting, and metal emission features. Furthermore, all these peculiarities are variable.

In this paper we analyze the manifestations of circumstellar envelopes in the optical spectra of PPNe paying special attention
to the stars whose atmospheres, according to previous studies, underwent evolutionary variations of the chemical composition.
Section~\ref{objects} briefly describes the employed observational data  and lists the studied stars and their basic
parameters. In addition to PPNe, we also study several related luminous stars with similar properties, including objects with
unclear evolutionary status. In Section~\ref{analys} we analyze the available data on the peculiarities of the profiles
of the H$\alpha$, Na\,I doublet, and metal lines, found in high-resolution spectra, as well as data on the presence of
molecular bands and outflow velocities for objects with different envelope structures. In Sections~\ref{discuss} 
and~\ref{conclus} we discuss the obtained results  and summarize the main conclusions.

\section{OBSERVATIONAL DATA}\label{objects}

Over the past decade AGB and post-AGB supergiants with IR excesses and several luminous stars with unclear evolutionary status have
been spectroscopically monitored with the 6-m telescope of the Special Astrophysical Observatory (V.G.~Klochkova). As a result,
a collection of high-quality spectra has been acquired with the primary purpose of searching for anomalies of stellar chemical
composition due to the nucleosynthesis of chemical elements in the interiors of low- and intermediate-mass stars (with masses smaller
than $8\div9 \mathcal{M}_{\odot}$) and the subsequent dredge-up of the synthesis products to the surface layers of stellar
atmospheres. These observational data are also used to search for peculiarities in the PPN spectra, to analyze the velocity fields
in the atmospheres and envelopes of these stars with mass loss, and to search for the likely long-term spectral and
velocity-field variations.

\newpage

\renewcommand{\baselinestretch}{0.9}

\begin{table}[]
\caption{List of supergiants of various types studied.  We adopt the types and spectral types of the objects from 
         the SIMBAD database.  The last column lists the references to  the papers reporting the main results for 
         the corresponding objects, including their effective temperatures.}  
\begin{tabular}{c|c|c|l|l|l}        
\hline
Star  & IRAS & Type of object  & Sp & $T_{\rm eff}$, K   & References\\
\hline
EM\,VES\,695       & 00470+6429       & emission    & B2--3   &     & [\cite{00470}]  \\
GSC\,04501--00166 & 01005+7910       & PPN         & B2\,Iab:e &21500 & [\cite{01005}] \\
XX\,Cam          &                   & R\,CrB$^1$  & G\,Iab:e  & 7250 & [\cite{XXCam}] \\
GSC\,02381--01014 & 04296+3429       & PPN         & G0\,Ia    & 6300 & [\cite{04296}]  \\
BD\,+48$\degr$1220& 05040+4820      & PPN         & A4\,Ia    & 7900 & [\cite{IBVS,05040}] \\
BD\,--6$\degr$1178& 05238--0626    & var?        &F5\,III\,+\,F3\,IV& & [\cite{05238}] \\
HD\,56126        & 07134+1005       & puls SR$^2$ & F5\,Iab   & 6600 & [\cite{56126,07134}]  \\
St\,H$\alpha$\,62& 07171+1823       & emission    & B0.5\,I   &21000 & [\cite{07171}] \\
AI\,CMi          & 07331+0021       & puls        &G5\,Iab  & 4500 & [\cite{AICMi}] \\
V510\,Pup        & 08005--2356     & puls SR     &F5\,Iae  & 7300 & [\cite{08005}]  \\
HD\,82040        & 09276+4454       & binary      &M6       & 3400 & [\cite{09276}] \\
LN\,Hya          & 12538--2611     & puls SR     &F3\,Ia     &      & [\cite{FHGL1,LNHya}] \\
Z\,UMi           & 15060+8315       & var Mira$^3$&         & 5250 & [\cite{ZUMi}] \\
R\,CrB           & 15645+2818       & var, carb   &G0\,Iab:pe&     & [\cite{RCrB}]   \\
UU\,Her          &                      & puls SR     & F2\,Ib  & 6200 & [\cite{UUHer1,UUHer2}]   \\
M12 K\,413       &                      &             &         & 4800 & [\cite{K413}]\\
M12 K\,307       &                      & W\,Vir$^4$  &         & 5600 & [\cite{K307}]   \\
V4334\,Sgr       &                      & nova-like   &F2-3\,II & 7250 & [\cite{sakurai}]  \\
V814\,Her        & 17436+5003       & puls SR     & F3\,Ib    & 7100 &  [\cite{FHGL2,19475}]  \\
89\,Her          & 17534+2603       & puls SR     & F2\,Ibe   &      &  [\cite{FHGL1}]  \\
V886\,Her        & 18062+2410       & Be          &B1\,IIIpe&      &  [\cite{18062}] \\
V887\,Her        & 18095+2704       & puls SR     &F3\,Ib   & 6500 &  [\cite{07134,Sahin}]   \\
GSC\,00439--00590 & 18123+0511       & PPN         & G5      & 4500 & [\cite{18123}]\\
R\,Sct           & 18448--0545     & RV\,Tau-type& K0\,Ibpv  & 4500 &  [\cite{RSct}] \\
HD\,179821       & 19114+0002       & puls SR     & G5\,Ia    & 5000 & [\cite{19114}]  \\
IRC\,+10420      & 19244+1115       & HG$^5$      & F8\,I--G0\,I& 9200 & [\cite{IRC1,IRC2}]  \\
HD\,331319       & 19475+3119       & PPN         & F3\,Ib    & 7200 & [\cite{19475}]\\
V5112\,Sgr       & 19500--1709     & PPN         & F2/F3\,Iab& 8000 &  [\cite{Reyn,19500}] \\
CGCS\,6857       & 20000+3239       & PPN         & G8\,Ia    & 5000 & [\cite{20000}]  \\
V1027\,Cyg       & 20004+2955       & puls        & G7\,Ia    & 5000 &  [\cite{20004}]  \\
QY\,Sge          & 20056+1834       & puls SR     & G0e     & 6250 & [\cite{QYSge}]  \\
FG\,Sge          & 20097+2010       & PPN         &B4\,Ieq--K2\,Ib  & 5500 & [\cite{FGSge1,FGSge2}]  \\
V1853\,Cyg       & 20462+3416       & puls        & B1\,Iae   & 20000& [\cite{V1853Cyg}]\\
GSC\,01655--00558 & 20508+2011       &             &         & 4800 &  [\cite{20508}]   \\
V2324\,Cyg       & 20572+4919       & PPN         & F0\,III   & 7500 & [\cite{20572}]    \\
V1610\,Cyg       &  RAFGL\,2688         & PPN         & F5\,Iae   & 6500 & [\cite{Egg1,Egg2}]  \\
V448\,Lac        & 22223+4327       & puls SR     & F9\,Ia    & 6500 & [\cite{Reyn,22223}]  \\
V354\,Lac        & 22272+5435       & puls var    & G5\,Ia    & 5600 & [\cite{22272,V354Lac,V354LacK}] \\
CGCS\,6918       & 23304+6147       & PPN         & G2\,Ia    & 5900 & [\cite{23304}] \\
$\rho$\,Cas      & 23518+5713       & HG$^5$      & G2\,Ia0e  & 5900 & [\cite{rhocas}] \\
\hline 
\multicolumn{6}{p{125mm}}{\footnotesize $^1$ R\,CrB type variable, $^2$ semiregular variable, $^3$ Mira,
          $^4$ W\,Vir type variable in a globular cluster, $^5$ yellow hypergiant.}  \\
\end{tabular}
\label{stars}             
\end{table}

Here we use the data acquired in the Nasmyth focus with the NES~[\cite{nes1,nes2}] and Lynx~[\cite{lynx}]
echelle spectrographs. The NES spectrograph, equipped with a 2048$\times$2048 CCD and an image slicer~[\cite{slicer}], 
produces a spectroscopic resolution of R$\approx60\,000$. Since 2011 the NES spectrograph has been equipped with a 2048$\times$4096 
CCD which made it possible to significantly extend the wavelength coverage. The Lynx spectrograph, equipped with a 1k$\times$1k CCD,
produces a spectroscopic resolution of R$\approx$25\,000. The spectra of the faintest program objects (the stars K\,307 and
K\,413 in the globular cluster M\,12, V1027\,Cyg, V4334\,Sgr, the optical components of the IR sources IRAS\,04296+3429, 18123+0511,
23304+6147, and other stars with apparent magnitudes $V\ge 13^{\rm m}$) were acquired with the PFES echelle spectrograph
mounted in the primary focus of the 6-m telescope~[\cite{pfes}]. 

This spectrograph, equipped with a 1k$\times$1k CCD, 
produces a spectroscopic resolution of R$\approx$15\,000. We described the details of spectrophotometric
and position measurements of the spectra in our earlier papers, the corresponding references can be found in
Table~\ref{stars}. 

Note that we had to modify substantially the standard {\tt ECHELLE} context of the MIDAS
system, because of the image slicer employed. We used the software
described in~[\cite{ECHELLE}] to extract the data from the two-dimensional echelle spectra.

\begin{table}[h]
\caption{Basic data for selected post-AGB stars}
\medskip
\begin{tabular}{c|c|c|l|l|l|l}
\hline
 Object  & 21~$\mu$m & [C/Fe]  & Morphology & Type of  & \multicolumn{2}{c}{$V_{\rm exp}$, km\,s$^{-1}$} \\
\cline{6-7}
         &        &  and [s/Fe]  & of the envelope & the C$_2$ bands   &   &   \\
         &        &             &  [\cite{Sahai,Ueta2000,Siodmiak,Lagadec}] &     &  CO     &  optics   \\
\hline
02229+6208& +      & +    &bipolar              &abs. [\cite{Reddy1999}]  & 10.7 [\cite{Hrivnak}]  &  8--15 [\cite{Reddy1999}]  \\[5pt]
04296+3429& +      & +    &bipolar\,+\,halo\,+  & emiss. [\cite{04296}]   & 10.8 [\cite{Hrivnak}]  &  3 [\cite{04296}]  \\%[-5pt]
          &        &      & bar                 & abs.   [\cite{Bakk97}]  &                        & \\ % [-5pt]
05113+1347& +      & +    & unresolved          & abs.   [\cite{Bakk97}]  & 8--10 [\cite{Loup}]    & 6.3 [\cite{Reddy}] \\ [5pt]
05341+0852& +      & +    &  elongated halo     & no     [\cite{Bakk97}]  &                        & 10.8 [\cite{Reddy1997}] \\  %[-5pt]
          &        &      &                     & abs.   [\cite{Reddy1997}]&                       &   \\ [5pt]
07134+1005&   +    & +    &elongated halo       & abs.   [\cite{Bakk97,atlas}]& 10.2 [\cite{Hrivnak}]& 11 [\cite{atlas}]  \\ [5pt]
08005$-$2356  &\multicolumn{2}{c|}{data unavailable} & bipolar& abs. [\cite{08005}]& 100: [~\cite{Hu}]& 42 [\cite{08005}]\\ %[-5pt]
              &\multicolumn{2}{c|}{}            &           & uncertain [\cite{Bakk97}]   &        &            \\ [5pt]
12538$-$2611&  $-$ &$-$  & unresolved           &  no [\cite{LNHya}]       &                          &      \\ [5pt]
19475+3119&  $-$   &$-$  &quadrupole\,+\,halo   & no  [\cite{19475}]       & 16.2, 20.1 [\cite{Hrivnak}] &      \\[5pt]
19500$-$1709  &   + & +  &bipolar               & no  [\cite{19500}]       & 17.2 and              & 20 and \\ %[-5pt]
              &     &    &                      &                          &   29.5 [\cite{Hrivnak}]&  30 [\cite{19500}]\\ [5pt]
20000+3239&   +    & +   &elongated halo        & abs. [\cite{Bakk97,20000}]&  12.0 [\cite{Omont}] & 11.1 [\cite{20000}] \\[5pt]
RAFGL\,2688   &   +& +   &multipolar\,+         & abs. [\cite{Bakk97}]      & 17.9, 19.7 [\cite{Loup}] & 22.8 [\cite{Bakk97}] \\ %[-5pt]
              &    &     & halo\,+\,arcs        & emiss. [\cite{Egg1}]      &                      & 60 [\cite{Egg1}]    \\ [5pt]
22223+4327&   +    & +   &halo+small lobes      & abs. [\cite{Bakk97}]      & 14--15 [\cite{Loup}] & 14.0 [\cite{Bakk97}] \\ %[-5pt]
              &        & &                      & emiss. [\cite{22223}]     &                      & 15.2 [\cite{22223}]  \\[5pt]
22272+5435&   +    & +   &elongated halo\,+\,arcs& abs. [\cite{V354LacK,Bakk97}] & 9.1--9.2 [\cite{Hrivnak}]& 10.8 [\cite{V354Lac}] \\%[-5pt]
              &        & &                      &                           &                      &  11.6 [\cite{Bakk97}] \\ [5pt]
23304+6147&   +    & +   &quadrupole\,+\,halo\,+& abs. [\cite{Bakk97}]      & 9.2--10.3 [\cite{Hrivnak}] &  15.5 [\cite{Bakk97}]  \\%[-5pt]
              &    &     &  arcs               & emiss. [\cite{23304}]      &                          & $\approx\,20$ [\cite{23304}]\\ [5pt]
\hline 
\multicolumn{7}{c}{ } \\

\multicolumn{7}{p{180mm}}{\it Carbon [C/Fe] and heavy-metal [s$/$Fe] overabundance (or lack thereof) in the atmosphere of the 
              central star is indicated by the ``+'' sign in the third column in accordance with the results of 
              Reddy et al.~[\cite{Reddy1999,Reddy,Reddy1997}] for IRAS\,02229+6208, 05113+1347, and 05341+0852 respectively;
              Klochkova et al.~[\cite{04296,07134,08005,19475,23304}] for IRAS\,04296+3429, 07134+1005, 08005$-$2356, 19475+3119, 
              and 23304+6147 respectively;  Klochkova~[\cite{19500}] for V5112\,Sgr; Kipper and Klochkova~[\cite{20000}] for                                                                                        
              IRAS\,20000+3239;  Klochkova et al.~[\cite{Egg2}] and Ishigaki et al.~[\cite{Ishigaki}] for RAFGL\,2688; Klochkova et al.~[\cite{22223}]
              Klochkova et al.~[\cite{V354Lac}] for V354\,Lac. The last column  gives the expansion velocity of the envelope 
              for V448\,Lac; as determined  from the position of C$_2$ Swan bands. The expansion velocity for  
              IRAS\,19500$-$1709 is determined from the envelope components of the Ba\,II lines.}  \\
\end{tabular}
\label{PPN}
\end{table}

\renewcommand{\baselinestretch}{1.0}

\section{PECULIARITIES OF THE OPTICAL SPECTRA OF POST-AGB STARS}\label{analys}

Our comprehensive study of the program stars allowed us to determine (or refine) their evolutionary status. One of the
results of our analysis is  that the studied sample of luminous stars with IR--excesses is not
homogeneous~[\cite{BTA-evolution}]. It follows from Table~\ref{stars} that the program objects  include
luminous stars of various types ranging from low-mass W\,Vir type variables to hypergiants. In this paper we analyze the
peculiarities of the optical spectra of  post--AGB stars paying special attention to the subsample of objects listed in
Table~\ref{PPN}. It contains  objects with central stars whose atmospheres are overabundant in carbon and heavy
metals and whose circumstellar envelopes have a complex morphology and are usually rich in carbon, as evidenced by  
the presence of C$_2$,  C$_3$, CN, CO, etc. molecular bands in their IR, radio,
and optical spectra.  Furthermore, the objects from Table~\ref{PPN} are among those few PPNe whose
IR-spectra exhibit the so far unidentified emission band at 21\,$\mu$m~[\cite{Kwok,Hrivnak2009}]. 
Despite an extensive search for appropriate chemical agents, so far no conclusive identification has been proposed 
for this rarely observed feature. However, its very presence in the spectra of PPNe with carbon enriched envelopes 
suggests that this emission may be due to the presence of a complex carbon-containing molecule
in the envelope (see~[\cite{Hrivnak2009,Li}] for details and references).

In addition to a sample of related objects with the above features,  Table~\ref{PPN} also includes the infrared
sources  \mbox{IRAS\,08005$-$2356}, 12538$-$2611, and 19475+3119. The object  IRAS\,08005$-$2356 has so far been 
poorly studied, and no data are available either on the peculiarities of the chemical composition of its
atmosphere or on the presence of the 21-$\mu$m band. However, IRAS\,08005$-$2356  can be viewed as related to the objects of
this sample because the optical spectrum of its central star V510\,Pup exhibits C$_2$~Swan bands, the hydrogen and 
metal lines in its spectrum have emission-absorption profiles~[\cite{08005}], and the circumstellar envelope
is observed in CO emission~\cite{Hu}. The infrared source IRAS\,12538$-$2611 is associated with the high-latitude
supergiant LN\,Hya, which  exhibits a number of spectral peculiarities~\cite{LNHya} deserving a detailed
discussion within the context considered. IRAS\,19475$+$3119 also has a number of exceptional peculiarities, e.g., 
a strong overabundance of helium in the atmosphere of its central star~[\cite{19475}].

The principal types of spectral features in the optical spectra of PPNe are:  
\begin{enumerate}
 \item{low-- or moderate--intensity metal  absorptions, whose symmetric profiles show no apparent distortions;}  
 \item{complex profiles of neutral hydrogen lines, which vary with time and include absorption and emission components;} 
 \item{the strongest metal absorptions with low  excitation potentials of the low level, their variable profiles are often distorted by envelope features
       that make the profile asymmetric or split it into several components;} 
 \item{absorption or emission bands of mostly carbon containing molecules;} 
 \item{envelope components of the  Na\,I and K\,I resonance lines; (6)~narrow permitted or  forbidden metal emission lines that form in envelopes.} 
\end{enumerate}

 The presence of type 2--6 features is the key difference  of the spectra of PPNe from those of massive supergiants.

Our sample of PPNe, thoroughly studied by high-resolution spectra (Table~\ref{PPN}), practically coincides with the list
of C--rich protoplanetary nebulae the photometric and spectral properties of which were extensively studied by Hrivnak et
al.~[\cite{Hrivnak2009,Hrivnak2010}]. Of fundamental importance to us is the conclusion of Hrivnak et
al.~[\cite{Hrivnak2011}] about the rare occurrence of binaries among post-AGB stars of the type considered, which is
based on the long-term investigation of the velocity field in PPN atmospheres conducted by the above authors.
Van~Winckel~[\cite{Winck2007}], on the contrary, finds binary occurrence to be very high among  RV\,Tau type variables
with near-IR excesses, which also undergo the post--AGB stage of evolution. Thus, the results of Hrivnak et 
al.~[\cite{Hrivnak2009,Hrivnak2010,Hrivnak2011}] provide further evidence for the homogeneity of the PPN sample
considered here.

\subsection{H$\alpha$ Line}

The H$\alpha$ lines in the spectra of PPNe have complex (they combine emission and absorption components) and variable profiles
of various types: profiles with an asymmetric P\,Cyg type core or inverse P\,Cyg type profiles with two emission components in the
wings. The profiles often combine the features of both types. The presence of emission features in the  H$\alpha$ line is indicative
of a high mass loss rate~[\cite{Trams}] and serves as an indicator used for the search and identification of PPNe (see,
e.g., a review by Kwok~[\cite{Kwok}]). A good guidance is provided by the atlas of the HD\,56126
spectra~[\cite{atlas}], which illustrates the main features of the optical PPN spectra mentioned above. The
combination of observed properties (a typical double-humped spectral energy distribution,  F--supergiant spectrum with a
variable absorption and emission  H$\alpha$ profile, C$_2$ Swan bands in the optical spectrum that form in the outflowing extended
envelope, high overabundance in carbon and s--process heavy metals dredged up to the surface layers of the atmosphere as a result of
mixing) makes this object  a canonical post--AGB star. The H$\alpha$ profile in the spectrum of this star exhibits a striking
degree of variability: it follows from Fig.\,2 in the atlas~[\cite{atlas}] that all the above profile
types---asymmetric core, direct or inverse P\,Cyg type profile, and profiles with two emissions in the wings---were observed at
different times during decade-long observations with the 6--m telescope.

The complex H$\alpha$ profile in the spectrum of  V448\,Lac consists of a narrow core and broad  wings~[\cite{22223}]. 
According to the classification of S\'anchez Contreras et al.~[\cite{Contreras}], this profile can be classified as 
belonging to the type ``EFA'' (emission--filled absorption). H$\alpha$  profiles of the same type
are observed in PPN spectra irrespective of whether or not the atmosphere of the central star is enriched in carbon and heavy
metals. For example, H$\alpha$ lines with EFA profiles have been recorded in the spectra of stars with enriched atmospheres
associated with the following sources: IRAS\,04296+3429~[\cite{04296}], 05341+0852~[\cite{Reddy1997}],
07134+1005~[\cite{atlas}], 19500$-$1709, 22223+4327~[\cite{22223}], and 23304+6147~[\cite{23304}]  Profiles of
the same type were also recorded in the spectra of objects with non-enriched stellar atmospheres:
IRAS\,05040+4820~[\cite{05040}], 18095+2704~[\cite{07134}], and 19475+3119~[\cite{19475}]. We can make 
a similar conclusion that H$\alpha$ lines with EFA profiles are observed in the spectra of objects with different 
envelope morphologies.

\begin{figure}
\includegraphics[angle=-90,width=0.6\columnwidth,bb=30 60 570 780,clip]{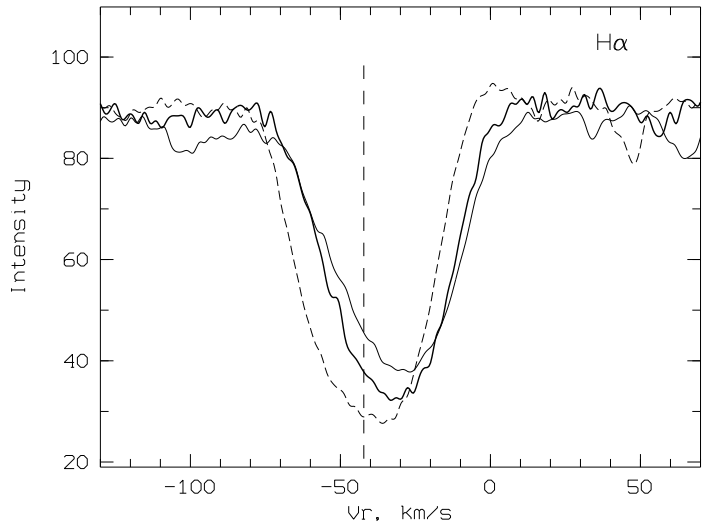}
\caption{Variability of the  core of the H$\alpha$ profile in the spectrum of V448\,Lac~\cite{22223}. 
         The vertical dashed line shows the systemic velocity. Here and in the
         subsequent figures showing parts of the spectra, the intensity
         along the vertical axis is plotted in relative units $I/I_{\rm cont}$ with the 
         continuum intensity assumed to be equal to 100 units.} 
\label{V448Lac-Halpha}
\end{figure}

Figure~\ref{V448Lac-Halpha} shows the H$\alpha$ core in the ``intensity--radial velocity ($V_r$)'' coordinates as
observed in the spectra of V448\,Lac taken in 2005 and 2008. This figure illustrates the variability of the  
H$\alpha$ core, and the radial velocity measurements~[\cite{22223}] indicate that the H$\alpha$ core is systematically 
blue-shifted relative to the metal lines. Klochkova et al.~[\cite{22223}] concluded that the profile of this line observed 
in the spectrum of  V448\,Lac is inconsistent with the theoretical LTE profile computed assuming normal (solar) hydrogen 
abundance. This discrepancy suggests that the envelope contributes to the formation of the profile and (or) that 
the H$\alpha$ formation conditions deviate from LTE. The available combination of the observed properties 
(effective temperature $T_{\rm eff}$, luminosity class, light variations, unstable behavior of atmospheric line 
$V_r$ variations, peculiarities of the optical and IR--spectra, including the variable peculiar profile of
the H$\alpha$ line, the asymmetry of the strongest absorptions, the combination of molecular bands, the similarity 
of the elemental abundance pattern in the atmosphere, and similar envelope morphology) is indicative of the analogy between
V448\,Lac  and HD\,56126.

\begin{figure}
\includegraphics[angle=0,width=0.6\columnwidth,bb=30 50 560 790,clip]{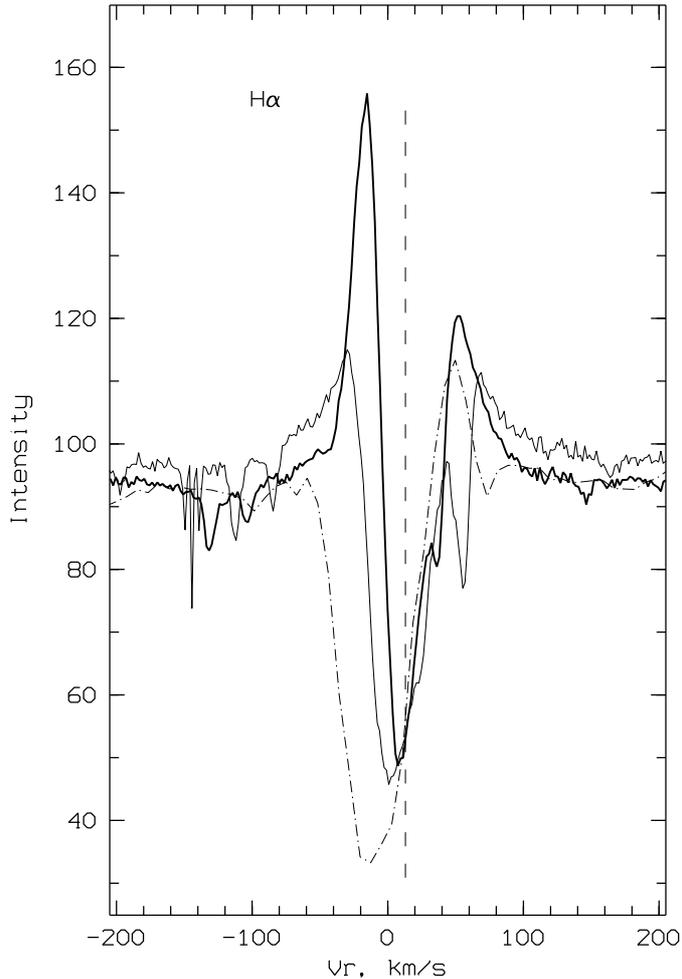} %V5112Sgr-Halpha
\caption{Variability of the absorption and emission profile of the H$\alpha$ line in the spectra of the high-latitude post--AGB
         supergiant V5112\,Sgr~[\cite{19500}] taken on different dates: August 2, 2012 (the solid thin line) and September 28, 2010 
         (the solid bold line). The dash-dotted line shows the profile in the spectrum  taken on July 5, 1996 with the LYNX
         spectrograph with a resolution of R\,=\,25\,000~[\cite{lynx}]. The vertical dashed line indicates the systemic velocity.}
\label{V5112Sgr-Halpha}
\end{figure}

A typical example of a profile with two emission components is the H$\alpha$  profile of the high-latitude post--AGB supergiant
V5112\,Sgr recorded at three observing epochs~[\cite{19500}] and shown in Fig.\,\ref{V5112Sgr-Halpha}. The spectrum of this
object, whose envelope even the Hubble Space Telescope failed to resolve~[\cite{Siodmiak}], exhibits variations of both
the intensities and positions of the absorption and emission components of the  H$\alpha$ line.

\begin{figure}
\includegraphics[angle=0,width=0.5\columnwidth,bb=30 50 560 780,clip]{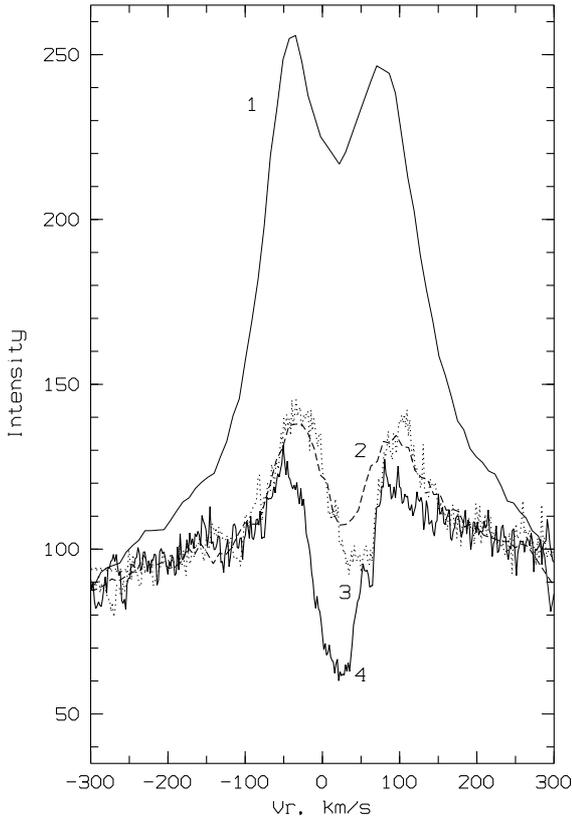} %IRAS20508-Halpha
\caption{Variability of the absorption and emission profile of the H$\alpha$ line in the spectrum of the 
         optical component of the IR--source IRAS\,20508+2011: observations made in 1999 (curve 1), 2000 (curve 2), 
         2003 (curve 3), and 2004 (curve 4)~[\cite{20508}].}
\label{IRAS20508-Halpha}
\end{figure}

The spectrum of the faint near-AGB optical component of IRAS\,20508+2011 shows a different type of profile and a different
spectral variability pattern. The powerful extended wings of the profile in Fig.~\ref{IRAS20508-Halpha} are typical for
such a cool supergiant ($T_{\rm eff}=4800$~K~[\cite{20508}]) at the close to AGB--stage.
The H$\alpha$ emission has a variable absorption superimposed on it. According to the classification of S\'anchez Contreras et
al.~[\cite{Contreras}], the H$\alpha$ profile of this object in our earliest spectra can be classified as type ``PE''
(pure emission). S\'anchez Contreras et al.~[\cite{Contreras}] recorded PE type profiles in the spectra of two AGB objects: 
IRAS\,09452+1330 and 10131+3049. The broad extent of the emission wings is indicative of a high
stellar-wind velocity near the AGB--stage. The extent of the H$\alpha$ wings has remained almost unchanged throughout the five
years of our observations of IRAS\,20508+2011, but nonstationary absorption is gradually becoming the dominating component.
Radial velocity measurements~[\cite{20508}] showed that the absorption in H$\alpha$ is systematically redshifted by about
10~km\,s$^{-1}$ relative to the photospheric lines (the position of the absorption in  H$\alpha$ is consistent with that of the
metal lines only on one observing date). The intensity of the blue-shifted emission exceeds that of the redshifted emission at
all observing epochs.

As is evident from Fig.\,\ref{IRAS20508-Halpha}, the H$\alpha$ profile in the spectrum  of IRAS\,20508+2011
changed in 1999--2003 from a powerful bell-like emission with a weak absorption in the core to a two--peaked
emission. The emission intensity systematically decreased, and in the spectrum taken in 2003 (profile 3 in
Fig.\,\ref{IRAS20508-Halpha}) the central absorption was below the continuum level. The width of the emission-line
wings also changed in the process. The observed H$\alpha$ profile can be interpreted as a superposition of two lines 
of different nature: a photospheric absorption and a strong emission spanning a wide range of velocities that forms 
in the extended circumstellar structure. The wings of the resulting emission profile extend out to the radial velocity 
of ${\pm 250}$~km\,s$^{-1}$.

\begin{figure}
\includegraphics[angle=0,width=0.5\columnwidth,bb=30 90 560 780,clip]{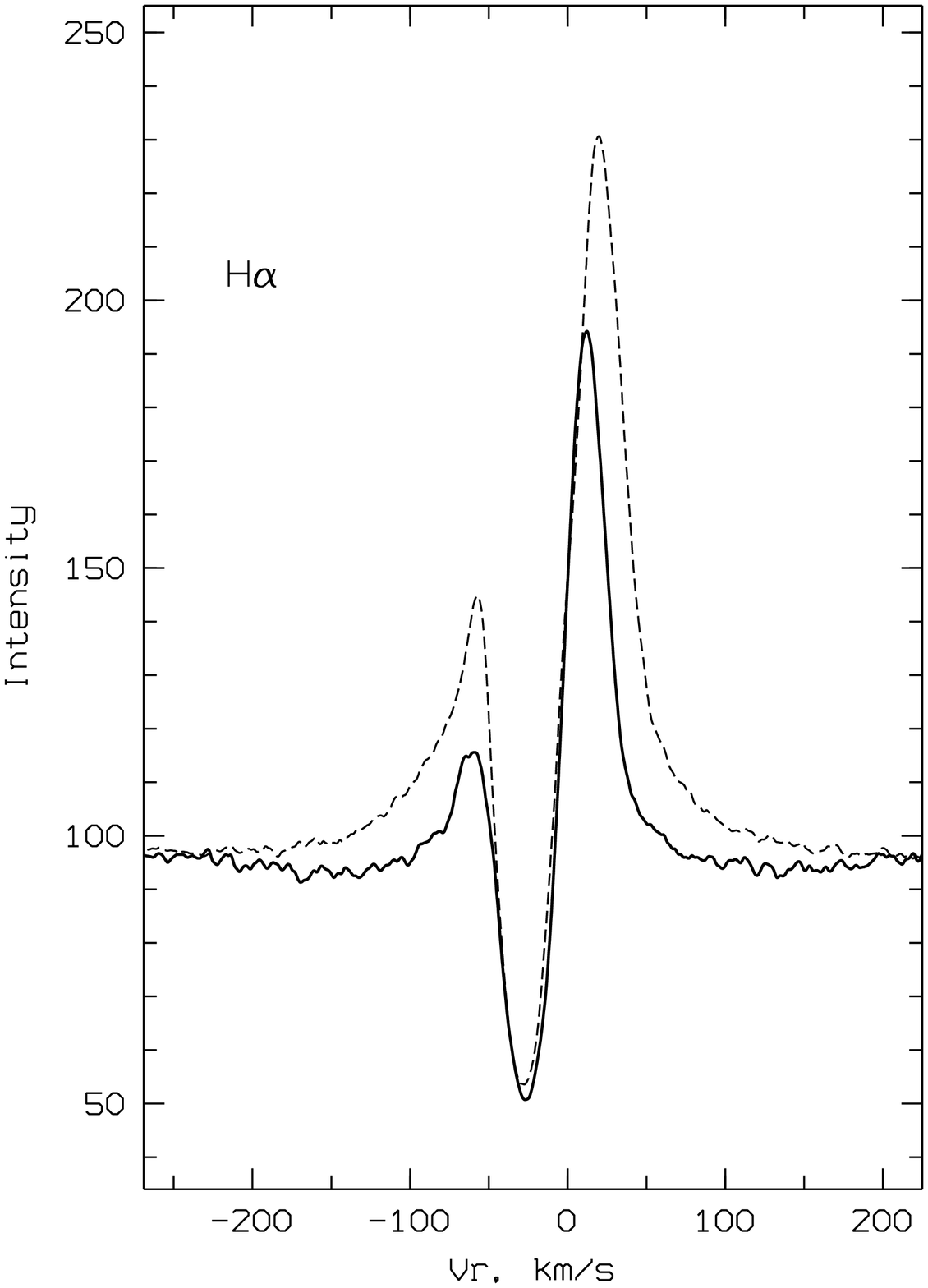} %IRAS05040-Halpha.ps
\caption{Variability of the absorption and emission profile of the H$\alpha$ line in the spectrum of  
         BD$\rm +48\degr 1220$, the optical component of the IR source IRAS\,05040+4820: March~8, 2004
        (the solid line) and January 10, 2004 (the dashed line)~[\cite{05040}].}
\label{IRAS05040-Halpha}
\end{figure}

The variability of the  H$\alpha$ emission is a well known phenomenon in AGB-- and post--AGB stars (see~[\cite{Oud,Klochkova1997}]
and references therein). The differences in the types of H$\alpha$ profiles and their variations are due 
to the instability of the dynamic processes in the extended atmospheres and envelopes of these stars: they can have the form
of a spherically symmetric outflow with constant or height dependent velocity, mass infall onto the photosphere, or 
pulsations. A two--component emission profile is indicative of a more complex, nonspherical wind structure, e.g.,
a circumstellar disk. An example of such a structure is the profile in the spectrum of   BD\,$\rm 48\degr1220=$\,IRAS\,05040+4820 shown
in Fig.\,\ref{IRAS05040-Halpha}: observations of this star made with the 6-m telescope resulted in the discovery of
its spectral variability and allowed detailed measurements of its chemical composition to be performed and its evolutionary
status to be determined~[\cite{05040}].

\subsection{Molecular Features}

Let us now consider the molecular component of the optical spectra of PPNe.
The spectra of protoplanetary nebulae with F--K-type supergiant central stars that have carbon-enriched atmospheres
show features of carbon-containing molecules  C$_2$, C$_3$,  CN, and CH$^+$. Kwok et al.~[\cite{Kwok-21}] report the
list of stars with 21~$\mu$m emission and C$_2$, C$_3$, and CN molecular bands in their spectra. Position 
measurements of molecular features in the spectra indicate that they form in expanding circumstellar
envelopes. Bakker et al. (see~[\cite{Bakk97}] and references therein) used high-resolution spectra to analyze 
a large sample of  post--AGB stars of this type.

We studied the PPN sample spanning a wide range of spectral types (temperatures) and identified C$_2$ Swan  bands 
in the spectra of some of these stars (see Table\,\ref{PPN}). The table also lists the expansion velocities 
corresponding to the band positions. The most thoroughly studied spectrum is that of the bright star HD\,56126~[\cite{atlas}]. 
In the 4010--8790~\AA\ wavelength interval, we performed a detailed identification and measured the positions of 
the rotational lines of several C$_2$ vibration bands, identified the CN and CH absorption bands, measured 
the depths and the corresponding radial velocities of numerous absorptions of neutral atoms and ions as
well as interstellar features. Besides the well-known variability of the  H$\alpha$ profile, we also found variations of the
profiles of a number of  Fe\,II and Ba\,II lines. The spectra taken in different years show no evidence of emission or
variability either in the C$_2$ molecular bands or in the Na\,I~D--lines. This fact is consistent with a rather 
simple elliptical shape of the nebula surrounding  HD\,56126.

It appears that the emission in the Swan bands or Na\,I D--lines is observed in the spectra of PPNe with bright 
and conspicuously asymmetric circumstellar nebulae. The results of spectroscopic observations of several PPNe 
confirm this hypothesis. An analysis of the spectra taken with the \mbox{6-m} telescope revealed  C$_2$ Swan 
emission bands of different intensities (relative to the continuum) in the spectra of the central stars of the
following sources: IRAS\,04296+3429~[\cite{04296}], IRAS\,08005$-$2356~[\cite{08005}],
RAFGL\,2688~[\cite{Egg1}], IRAS\,22223+4327~[\cite{22223}], and  IRAS\,23304+6147~[\cite{23304}]. According to HST
images~[\cite{Ueta2000,Siodmiak}], these objects have asymmetric  (and often bipolar) envelopes; this is
also evident from the polarization of their optical emission.

\begin{figure}
\includegraphics[angle=0,width=0.6\columnwidth,bb=50 110 570 780,clip]{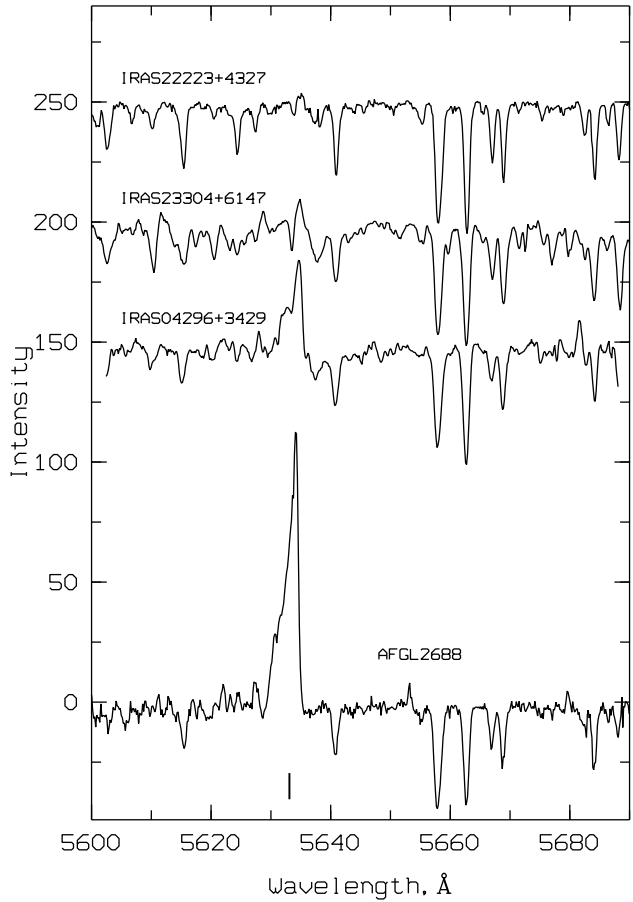} %Swan
\caption{A fragment of the spectra of selected PPNe from Table~2 containing the (0;1)  C$_2$ Swan band.}
\label{Swan}
\end{figure}

Figure\,\ref{Swan} shows fragments of the spectra of a number of objects containing  the (0;1)  C$_2$ Swan band.
As is evident from the figure, the Swan band is most intense in the spectrum of  RAFGL\,2688, which belongs to
a group of systems whose central parts (the star and the inner regions of the circumstellar envelope) undergo 
strong absorption in the gas and dust torus (or disk) and where the emission of the central parts
is scattered by the dust particles of the bipolar structure. RAFGL\,2688~[\cite{Egg1}] was observed at the 6--m
telescope with the spectrograph slit projected onto the northern lobe of the nebula. Because of such an arrangement 
combined with the strong absorption of the radiation of the central star, the
observed spectrum is dominated by the contribution of the emission of the envelope.

Klochkova et al.~[\cite{04296}] studied the spectrum of IRAS\,04296+3429, whose properties are similar to those of
RAFGL\,2688, and showed that the intensity ratio of different Swan bands corresponds to the resonance fluorescence mechanism.

\subsection{Lines of the  Na\,I Resonance Doublet and the Diffuse Bands}

In addition to the variable emission components of neutral hydrogen lines and molecular bands, which 
distinguish the spectra of PPNe from the spectra of normal supergiants, some of them contain features 
that form in the circumstellar medium.  Circumstellar absorption components of the  Na\,I and K\,I
resonance lines are a common feature found in the spectra of PPNe, as expected for objects with circumstellar absorption.
Measurements of the positions of circumstellar components are widely used to determine the envelope expansion velocities. 
This, in particular, applies to  HD\,161796, HD\,101584, and V354\,Lac, in whose spectra Bakker et al.~[\cite{Bakk1996}], 
Reddy et al.~[\cite{Reddy}], and Kipper~[\cite{Kipper2005,Kipper2007}] found circumstellar components of the 
Na\,I D--lines. Klochkova et al.~[\cite{V354Lac}] and Klochkova~[\cite{V354LacK}] used the spectra obtained
with the 6--m telescope to study the  Na\,I D--line profiles and other features in the spectrum of  V354\,Lac  in more detail. 
The circumstellar absorptions of the Na\,I~D--lines were also identified in the well-studied spectrum of
HD\,56126~[\cite{Bakk97,atlas,56126}]. Much more rarely the circumstellar envelope manifests itself in the form 
of emission components of the Na\,I D--lines. Examples include the spectra of  V510\,Pup, which is the optical
counterpart of the IR--source IRAS\,08005$-$2356~[\cite{08005}], bipolar nebula
RAFGL\,2688~[\cite{Egg2}], and semiregular variables QYSge\,=\,IRAS\,20056+1834~[\cite{Rao,QYSge}]
and 89\,Her~[\cite{Kipper2011}].

\begin{figure}
\includegraphics[angle=0,width=0.5\columnwidth,bb=30 60 570 780,clip]{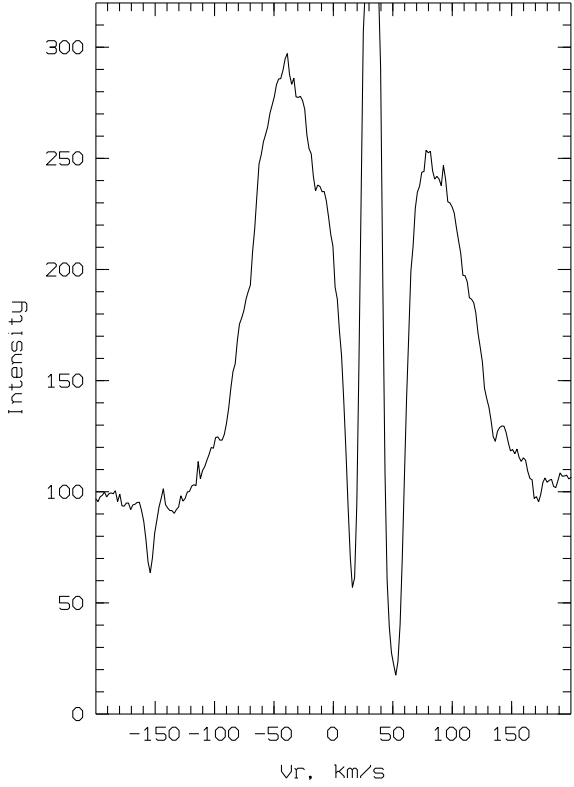} %QYSge-NaI
\caption{Profile of the Na\,I  D$_2$ line in the spectrum of QY\,Sge.} 
\label{QYSge-Na}
\end{figure}

We now discuss in more detail the spectrum of the yellow supergiant QY\,Sge, which belongs to a group of objects where the
radiation of the central star is obscured by a powerful envelope. Rao et al.~[\cite{Rao}] identified the main components  
(an absorption spectrum of a G--type supergiant combined with narrow low excitation emissions  and a powerful broad emission 
in the Na\,I D lines) in the spectrum of QY\,Sge and proposed for this object a model with a circumstellar torus and a bipolar mass
outflow. According to this model, the central star is completely obscured from the observer, and as a result we see the radiation
reflected from the inner wall of the torus. Multiple spectroscopic observations of QY\,Sge made with the 6--m telescope~[\cite{QYSge}] 
allowed the kinematic pattern in the atmosphere and envelope of this star to be studied in detail. As is evident from 
Fig.\,\ref{QYSge-Na}, the complex emission and absorption profile of the Na\,I~D$_2$ line in the spectrum of QY\,Sge 
includes a very broad emission component extending from $-170$ to $+120$~km\,s$^{-1}$. The central part of the broad emission 
is cut by an absorption feature, which, in turn, is split in two by a narrow emission peak (16~km\,s$^{-1}$ at r\,=\,2.5). 
The positions of the Na\,I~D emission features remain unchanged, and this fact indicates that the features form in regions 
that are external to the photosphere of the supergiant. 

The pattern of the profile variations of the emission and absorption lines and radial velocities as measured from different 
profile features is consistent with the model of a toroidal dust envelope that obscures the central source and bipolar cones 
filled with high-velocity gas. Both in the emission and absorption features of the spectrum, details have been identified that 
are indicative of spatial and temporal inhomogeneities in the dust and gas components of the object.

A possible relation between the type of the circumstellar nebula and the presence of a Na\,I D--emission component is difficult 
to find by analyzing a limited sample of objects.
It may be suggested that Na\,I D--emission is observed in the case where the extended envelope has a complex (for example -- bipolar) 
structure, because the three infrared sources IRAS\,08005$-$2356, RAFGL\,2688, and  IRAS\,20056+1834 with bipolar outflows exhibit 
Na\,I~D--emission in their spectra. However, contrary to this hypothesis, this emission also appears in the spectrum
of 89\,Her, where no circumstellar dust envelope has been observed so far~[\cite{Siodmiak}].

The so-called diffuse interstellar bands (DIBs) have been studied extensively for several decades since their discovery in the
1930s. The origin and chemical identification of these features remain a mystery despite numerous observations and laboratory
experiments (see, e.g., the review~[\cite{Sarre}]).

The correlation between the intensity of these bands and interstellar reddening (see~[\cite{Kos}] and references
therein) suggests that dust particles should play a certain part in the formation of DIBs. The similarity of the physical 
and chemical conditions in the interstellar medium and in circumstellar gas and dust envelopes led the researchers to look
for \mbox{DIB-like} features in the spectra of PPNe. As far back as 1993, Bertre and Lequeux~[\cite{Lequeux}] identified
similar features that formed in circumstellar envelopes in low-resolution spectra of some mass-losing stars. Zacs et
al.~[\cite{Zacs1999}] analyzed high-resolution spectra that we took with the 6--m telescope and suspected that
circumstellar components of DIBs may be present in the spectra of HD\,179821 and  V354\,Lac. The medium-resolution spectra of
HD\,331319 and the optical component of IRAS\,23304+6147 taken with the PFES spectrograph~[\cite{pfes}] also contain
strong absorptions whose positions possibly indicate  their circumstellar origin~[\cite{19475,23304}].
However, the spectroscopic resolution was not high enough to allow more definitive conclusions. Subsequent higher-resolution 
(R$\ge 60\,000$) observations made with the NES spectrograph mounted on the 6--m telescope revealed several features 
in the spectrum of V354\,Lac that could be identified with circumstellar bands~[\cite{V354Lac}]. The mean radial velocity
averaged over these features agrees, within the errors, with that determined from the circumstellar component of the 
profile of the Na\,I~D--lines. Such a coincidence may serve as evidence for the reality of circumstellar DIB analogs.

\begin{figure}
\includegraphics[angle=-90,width=0.6\textwidth,bb=30 70 570 780,clip=]{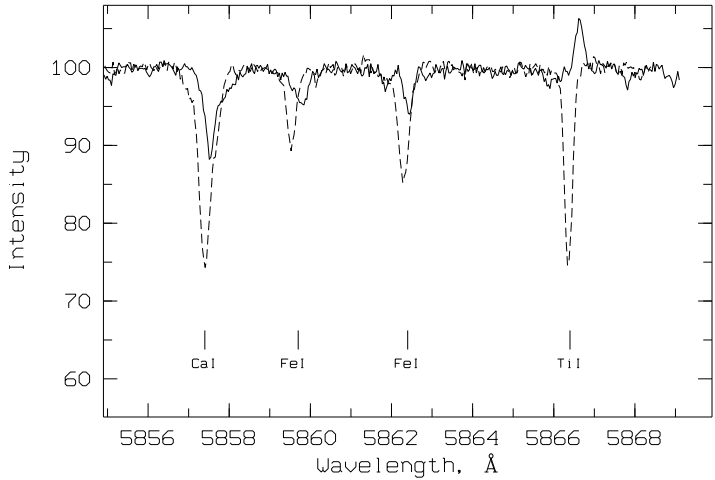} %LNHya-Ti5866.ps
\caption{A fragment with the  Ti\,I~5866.40~\AA{} emission in the spectra of LN\,Hya taken when the stellar atmosphere 
         was in the quiet (April 2, 2010, the dashed line) and excited (June 1, 2010, the solid line)  states. 
         Also shown are the identifications of the main spectral features that appear in this fragment.}
\label{LNHya-Ti}
\end{figure}

Luna et al.~[\cite{Luna}] searched systematically for DIBs in high resolution spectra of a sample of  post--AGB stars and
concluded that diffuse bands in the spectra of such stars arise in the interstellar medium, whereas the physical conditions in the
circumstellar envelopes of these objects do not favor the formation of these features. However, Klochkova~[\cite{19500}] 
recently showed that the optical spectrum of the high latitude  post--AGB supergiant V5112\,Sgr contains weak absorptions 
whose positions are indicative of their formation in the circumstellar envelope. The velocities V$_r$(DB) were measured 
in the 5780--6613~\AA\ wavelength interval from the positions of reliably identified 5780,  5797, 6196, 6234, and 6379~\AA\ 
features. The mean V$_r$(DB) averaged over several spectra and determined with an accuracy better than $\pm 0.5$~km\,s$^{-1}$  
agrees excellently with the velocity determined from the blue component of the Na\,I D--lines. The resulting agreement 
leads us to conclude that the weak bands found in the spectrum of V5112\,Sgr origin in the
circumstellar envelope. Kipper~[\cite{Kipper2013}] independently came to similar conclusions for the same object
based on the spectra of V5112\,Sgr taken with a different instrument.

These results mark a new stage in the search for diffuse circumstellar bands.

\subsection{Low-Excitation Emissions of Circumstellar Envelopes}

The spectra of selected PPNe were found to contain narrow emissions identified with low-excitation metal lines. For example,
Bakker et al.~[\cite{Bakk1996}] thoroughly studied the optical spectrum of the supergiant  HD\,101584 and found not only
symmetric high excitation absorptions and complex  P\,Cyg type profiles but also pure metal emissions, whose positions in the
spectrum match the systematic velocity and indicate that the corresponding features origin  in the stellar envelope.
Kipper~[\cite{Kipper2011}] also found numerous envelope emissions in the spectrum of 89\,Her.

\begin{figure}
\includegraphics[angle=-90,width=0.8\textwidth,bb=30 40 560 800,clip=]{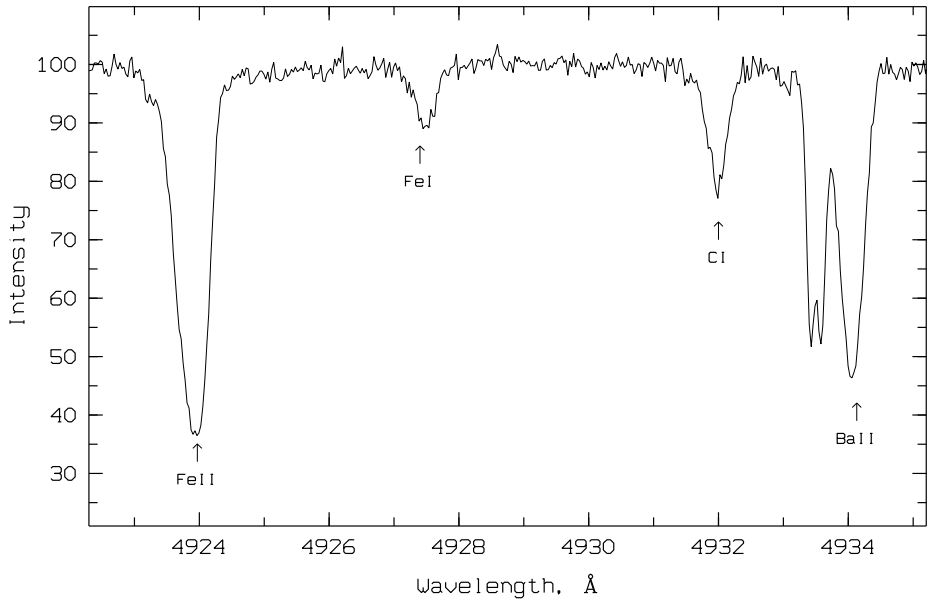} %V5112Sgr-Ba4934b
\caption{A fragment of the V5112\,Sgr spectrum taken on July 7, 2001, containing a split Ba\,II~4934~\AA\ line.}
\label{V5112Sgr-Ba4934}
\end{figure}

The spectrum of LN\,Hya taken on June 1, 2010 exhibits a peculiarity previously unknown for this star: weak emissions of
neutral atoms (V\,I, Mn\,I, Co\,I, Ni\,I, Fe\,I) with intensities amounting to several percent of the continuum
level~[\cite{LNHya}]. These emission features were absent in the earlier spectra taken with the 6--m telescope
before June 2000. Instead, rather strong absorptions corresponding to the same atomic transitions were observed.
Figure\,\ref{LNHya-Ti} shows, by way of an example, a fragment of the spectrum of LN\,Hya containing one of these
emissions---Ti\,I~5866.40~\AA, which appeared on June 1, 2010, when the stellar atmosphere was in an excited state. 
It follows from the figure that the spectrum contains the  Ti\,I~5866.40~\AA\
emission with an intensity of  about 6\%, whereas during the  period of quiet atmosphere (April 2, 2010) an absorption is
observed, whose position matches the radial velocity determined from other undistorted absorptions in the spectrum studied. The
intensity of the emissions decreases in the subsequent spectra taken in 2011, and some of these emissions disappear
altogether~\cite{LNHya}.  The atmosphere of the star returned to its normal state, and  in 2011 its spectrum, including
the  H$\alpha$ profile, was practically indistinguishable from the spectrum taken on February 21, 2003. Klochkova and
Panchuk~[\cite{LNHya}] could not simultaneously monitor the physical conditions in the stellar atmosphere and in the
envelope, and therefore could only suggest that the instability in the upper layers of the atmosphere of LN\,Hya developed, rapidly
amplified, and disappeared during 60 days from April to June 2010.

The constant average velocity of V$_{\rm sys}=-21.6$~km\,s$^{-1}$, measured from weak symmetric emissions of neutral 
atoms (V\,I, Mn\,I, Co\,I, Ni\,I, Fe\,I), can be adopted as the systemic velocity for LN\,Hya. The same velocity value 
was measured from the emission component of the Na\,I D lines, whose position also does not change with time.

Note that the stars mentioned in this section (89\,Her, LN\,Hya, and HD\,101584) constitute a  group of related objects 
in terms of chemical composition: they have low metallicity and are underabundant in heavy metals.

\subsection{Peculiar Profiles of Strong Absorptions}

The systematic monitoring of PPN candidates and related objects with envelopes performed with a high spectral resolution produced
a new result---the discovery of peculiar profiles of the strongest absorptions in the spectra of selected objects:
HD\,56126~[\cite{atlas}], V354\,Lac~[\cite{22272}], V448\,Lac~[\cite{22223}], and V5112\,Sgr~[\cite{19500}]. 
Let us illustrate this effect using the example of the spectrum of the supergiant  V5112\,Sgr, where it is most pronounced.

In the spectra of  V5112\,Sgr, taken in a wide wavelength interval, we found in addition to the variable H$\alpha$ profile
other, previously unknown peculiarities. The strongest absorptions have anomalous profiles: they are either asymmetric with 
extended blue wings or split into individual components.
Figure\,\ref{V5112Sgr-Ba4934} shows a fragment of the spectrum of  V5112\,Sgr taken on July~7, 2001 with a split
Ba\,II~4934~\AA\ line. The different widths of the components are immediately apparent: the red component is about twice 
broader than the blue components, which are offset substantially relative to the systemic velocity. This difference between 
the component widths indicates that the red and blue components form under different physical conditions. At the same time, 
the profiles of strong absorptions of iron-group metals in the spectrum of this star are neither asymmetric nor split, 
as is immediately apparent from the same Fig.\,\ref{V5112Sgr-Ba4934}, where we see a very strong but unsplit  Fe\,II~4924~\AA\ line.

\begin{figure}
\includegraphics[angle=0,width=0.5\columnwidth,bb=30 50 560 780,clip=]{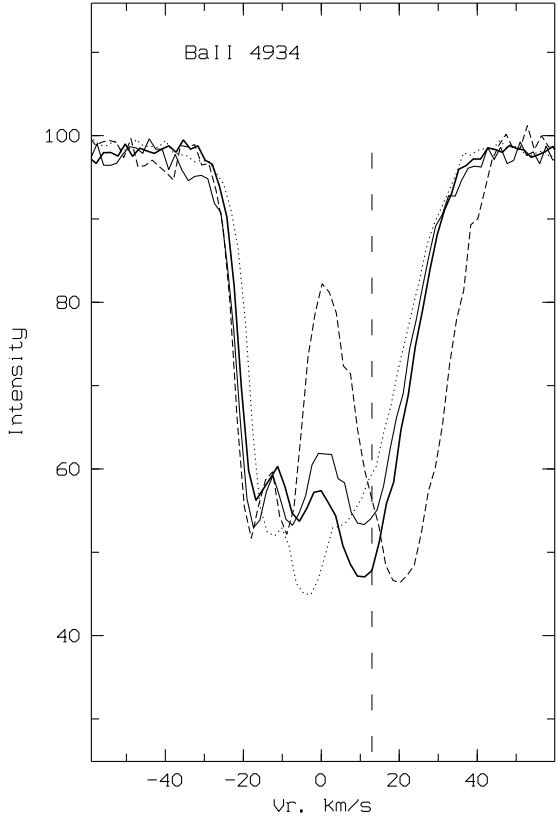} %V5112Sgr-Ba4934-prof
\caption{Variability of the  Ba\,II~4934~\AA\ line profile in the spectra of V5112\,Sgr taken in different years: 
         August 2, 2012 (the thin solid line); June 13, 2011 (the solid bold line); August 14, 2006 (the dotted line); 
         July 7, 2001 (the dashed line)~[\cite{19500}].} 
\label{Ba4934-prof}
\end{figure}

A comparison of the line profiles in the spectra of V5112\,Sgr taken during different nights reveals substantial variability of
the profile shapes and of the positions of the components of the split lines. To illustrate the variability effect, we show in
Fig.\,\ref{Ba4934-prof} the  Ba\,II~4934~\AA\ line profile, which is most asymmetric and most variable. It follows
from this figure that the position of the photospheric (red) component of the complex profile is variable, whereas the blue
components, which, as shown in~[\cite{19500}], form in the envelope, are stable.

The semiregular variable V354\,Lac is the closest analog to V5112\,Sgr among the objects listed in Table\,\ref{PPN}
in terms of the structure of the envelope and spectral peculiarities. The spectroscopic monitoring of this star~[\cite{V354Lac}] 
carried out at the SAO 6--m telescope with a resolution of $R=60\,000$ revealed the splitting of the strongest absorptions 
with the low-level excitation potential of  $\chi_{\rm low}\le 1$~eV.  An analysis of the kinematic pattern showed that 
the blue component of the split line forms in the powerful gas and dust envelope of V354\,Lac.  This splitting shows up most 
conspicuously in the profile of the strong Ba\,II~6141~\AA\ line. The shift of the blue component of the Ba\,II line 
coincides with that of the circumstellar component of Na\,I D lines, which forms in the same layers as the circumstellar
C$_2$ Swan bands. This coincidence indicates that the complex profile of the Ba\,II~6141~\AA\ line contains, in addition to the
atmospheric component, a component that forms in the circumstellar envelope. Such splitting (or the profile asymmetry  due to the
more shallow slope of the blue wing) is also observed for other Ba\,II ($\lambda\lambda$\,4554, 5853, 6496~\AA) lines as well as
for the strong Y\,II~5402~\AA{}, La\,II~6390~\AA, and Nd\,II~5234, 5293~\AA\ lines. The lines of these ions in the spectrum of
V354\,Lac are enhanced to the extent that their intensities are comparable to those of neutral-hydrogen lines. For example, the
equivalent width of the Ba\,II~6141~\AA\ line reaches $W_{\lambda}\approx1$~\AA, and that of  H$\beta$ reaches
$W_{\lambda}\approx2.5$~\AA.

The difference between the peculiarity types of the profiles of two pairs of stars---V5112\,Sgr and V354\,Lac with split profiles
of the strongest absorptions of selected elements, and HD\,56126 and V448\,Lac with asymmetric but unsplit profiles---suggests that
the morphology of the circumstellar envelope may be the factor that causes the peculiarity and variability of the profiles of the
strongest lines. As is evident from Table\,\ref{PPN},  the two stars V5112\,Sgr  and V354\,Lac with split absorptions
have bipolar envelopes, whereas the absorptions are unsplit in the spectra of HD\,56126 and V448\,Lac and the envelopes of these
stars have a less defined structure. 

This hypothesis is further corroborated by the three-component structure of the strong absorption profiles observed in 
the spectrum of   V5112\,Sgr,  where CO observations show both the slow (V$_{\rm exp}=10$~km\,s$^{-1}$) and the fast 
($30$--$40$~km\,s$^{-1}$) expansion~[\cite{Bujar}]. The profiles of the split lines include a photospheric component and 
two envelope components, one of which, like in the case of the CO profile, arises in the envelope that formed at the AGB--stage 
and expands at a velocity of V$_{\rm exp}(2)\approx 20$~km\,s$^{-1}$, and the other one arises in the envelope that moves 
at a velocity of V$_{\rm exp}(1)\approx 30$~km\,s$^{-1}$ and formed later. Similar peculiarities of the profiles of low 
excitation absorptions, the variability of the optical spectra, and the Swan emission bands were also found for two other objects from
Table\,\ref{PPN}: IRAS\,22223+4327~[\cite{22223}] and IRAS\,23304+6147~[\cite{23304b}].

The splitting of strong low excitation absorptions is observed in the spectra of luminous stars of various types. One of the
well-known cases is the complex kinematic pattern in the outflowing atmosphere of the yellow supergiant $\rho$\,Cas.
Current results of spectroscopic monitoring are reported in~[\cite{Lobel,rhocas}]. The splitting of absorptions at certain 
light-curve phases in the spectra of W\,Vir type variables (see Sanford's results~[\cite{Sanford}]for W\,Vir), 
classical Cepheids~[\cite{Kraft}], RV\,Tau type variables~[\cite{Preston}], and Miras has been known since the mid-20th 
century. Kovtykh et al.~[\cite{Kovtyukh2011}] studied the variation of the profiles of emission-containing lines in the 
spectrum of W\,Vir with pulsation phase and proposed a scheme based on two shocks localized in the atmosphere and envelope. 
A hydrodynamic model of the extended atmosphere of this star has been developed to explain these variations. Advanced spectroscopy 
tools have also made it possible to detect the doubling of metal lines in the spectra of
the short-period pulsating star RR\,Lyr~[\cite{Chadid}].

Sanford~[\cite{Sanford}] attributed the splitting of absorptions and the presence of strong emission in hydrogen and
helium lines in the spectra of pulsating stars to shock propagation  (the Schwartzschild
mechanism~[\cite{Schwarz}]). However, the efficiency of the Schwartzschild mechanism in the case of PPNe is open to
question, because atmospheric pulsations in the atmospheres of these objects have low amplitudes, 
$\Delta{\rm V_r} \approx 1\div 3$~km\,s$^{-1}$~[\cite{atlas,22223,V354Lac,V354LacK,Hrivnak2011}]. It therefore 
remains unclear how the shocks,  whose generation requires higher pulsation amplitudes  
($\Delta {\rm V_r}\ge 5\div10$~km\,s$^{-1}$~]\cite{Rao2005}]), may develop.

\section{DISCUSSION OF THE RESULTS}\label{discuss}

The primary conclusion of the analysis of the properties of the supergiants with infrared excess studied so far is that the
available sample of these objects is inhomogeneous. These objects are found  to exhibit a great variety of peculiarities in the
optical spectra of their central stars, the chemical composition of their atmospheres and envelopes, and the morphology and
kinematic state of their circumstellar envelopes. We should point out, in particular, that supergiants with infrared excess include
RV\,Tau type variables. These variable stars with near-infrared excesses undergo the  post--AGB evolutionary stage, and most of
them are binaries~[\cite{Winck2007}]. They owe their peculiar chemical composition to the high efficiency of selective
separation processes~[\cite{Giridhar2005}].

One of the important results of the study is the formation of a subsample of PPNe with atmospheres enriched in carbon and heavy
metals synthesized during the AGB evolution. It follows from Table\,\ref{PPN} that all these PPNe are also
distinctive in that their spectra contain the 21\,$\mu$m emission, which arises  in the circumstellar medium. Attempts to find an
interconnection between the morphology of the envelope and the peculiarities of the chemical composition of the central star
revealed no strict correlation. It may nevertheless be concluded that objects with enriched atmospheres (IRAS\,02229+6208,
04296+3429, 19500--1709, and 23304+6147) mostly have  structured (bipolar) envelopes. However, objects with enriched
atmospheres also include IRAS\,05113+1347, whose  envelope  even the HST failed to resolve, as well
as  IRAS\,05341+0852 and 07134+1005 with envelopes in the form of elongated haloes. One should also pay attention to
IRAS\,19475+3119, which has a structured envelope, whereas the atmosphere of its central star HD\,331319 is not enriched in
either carbon or heavy metals. However, helium lines were found and reliably measured in the spectrum of HD\,331319. 
The  presence of these lines in the spectrum of a star with T$_{{\rm eff}}=7200$\,K results in a significant helium 
overabundance in its atmosphere~[\cite{19475}], which can be viewed as a manifestation of helium synthesis during prior evolution.
 Helium overabundances were earlier found in HD\,44179\,$=$\,IRAS\,06176--1036~[\cite{Waelkens1992}]
and V5112\,Sgr~[\cite{Winck1996}].

The  H$\alpha$ lines in the spectra of PPNe usually  have complex profiles (a combination of absorption and emission
components) and are variable in time. It is evident that the surface-most atmospheric layers, which are susceptible to
stellar-wind influence, and the outflowing circumstellar envelope effect the formation of neutral hydrogen line profiles. 
We do not model these profiles in this paper, that is the aim of a separate study. Here we only emphasize the importance 
of spectroscopic monitoring in solving this problem.  A good example is the sharp change of the  H$\alpha$ profile 
in the spectrum of V5112\,Sgr which happened at the turn of the century:  in 1996 the line had the form of an inverse 
P\,Cyg type profile, and in the 2000s the profile already contained two emission components~[\cite{19500}]. 
Note  that this change of the H$\alpha$ profile type, which has been recorded in the spectra taken with the 6--m telescope, 
occurred simultaneously with a significant change of the photometric properties of the object as detected by Hrivnak 
et al.~[\cite{Hrivnak2010}] during their long-term monitoring.

The optical spectra of PPNe with enriched atmospheres are often found to exhibit spectral features of carbon-containing molecules.
In particular, the wavelength interval accessible to our instruments contains Swan C$_2$ bands. The data provided in
Table\,\ref{PPN} leads us to conclude that Swan bands in the spectra of PPNe are observed over a rather broad range of
effective temperatures of the central star, starting from the late G--type stars (T$_{\rm eff} \approx 5000$\,K). The spectra
of hotter objects with T$_{\rm eff}\ge 7500$\,K contain no molecular features because the molecules are destroyed by the
radiation of the hot star. The hottest star with Swan bands in our list is the source IRAS\,08005$-$2356, whose central star
has a temperature of T$_{\rm eff}= 7300$\,K. The presence of Swan bands in the cases where the central star is rather hot can
most likely be explained by the presence of two envelopes made of cool and hot dust. Bakker et al.~[\cite{Bakk97}]
determined the dust temperature in the envelopes of IRAS\,08005$-$2356 to be T$_{\rm dust}\le 150$\,K and
T$_{\rm dust}\ge 1200$\,K. According to the data of the above authors, Swan C$_2$ bands are observed in the spectra of
objects with cool dust envelopes (T$_{\rm dust}\le 300$\,K).

Note yet another peculiarity in the appearance of Swan bands in the spectra of the objects studied. It can be seen from
Table\,\ref{PPN} that the data on the presence or absence and the type of circumstellar Swan C$_2$ bands differ for
different observations of the same object. A comparison of the data on the presence of Swan bands according to Table\,3
in~[\cite{Bakk97}] with our data listed in Table\,\ref{PPN} leads us to conclude that discrepant
results are obtained for  common objects. For example, Bakker et al.~[\cite{Bakk97}] point out the 5635~\AA\ absorption
band in the spectrum of  RAFGL\,2688, where a strong emission band at 5635~\AA\ was found, according to the observations made with
the 6--m telescope (Fig.\,\ref{Swan}). In their recent study of the spectrum of RAFGL\,2688, Ishigaki et al.~[\cite{Ishigaki}]  
point out a very strong emission in the molecular bands. The spectra of IRAS\,08005$-$2356 taken
with the 6--m telescope~[\cite{08005}] show strong Swan absorption bands, whereas Bakker et al.~[\cite{Bakk97}]  report uncertain 
results for this object. Bakker et al.~[\cite{Bakk97}] report Swan absorption bands for the IRAS\,22223+4327 source with 
a powerful structured envelope, whereas more recent observations of this object made with the 6--m telescope show emission 
bands~[\cite{22223}].

These discrepancies may be due both to temporal variations of the physical conditions in the circumstellar envelopes of PPNe (caused
by the deviation from symmetry and sphericity) and different observing conditions for extended structures when studied with
different telescopes: the differences in the  spectrograph slit orientation and seeing may result in different flux contributions
from the central star and the envelope. Note, for example, a significant difference between the measured intensities of the
emission component of the Na\,I D--ines reported by Klochkova et al.~[\cite{Egg1}] and Ishigaki et al.~[\cite{Ishigaki}]. 
The large difference between the measured fluxes is due not only to the factor-of-two difference in the spectral resolution of 
the data reported in~[\cite{Egg1}] and~[\cite{Ishigaki}] but also to physical factors. It is evident that there may be objects
with such dust disk orientation that the central star is not obscured from the observer, and its flux is not reduced by several
factors of ten. In this case both the torus (or disk) and the scattering lobes are difficult to detect optically because of
their proximity to the bright star, and no bipolar nebula is observed. The spectrum of the central star in a system with such
orientation may show no molecular emission bands, because the emission in these bands is  ``drowned'' in the radiation of the
photosphere. The difference between the morphologies of ``extended halo'' and ``bipolar'' type nebulae may be purely visual, because
the observed shape depends strongly on the inclination of the structure to the line of sight and the angular
resolution~[\cite{Sahai}]. For example, according to the data of Nakashima et al.~[\cite{Nakashima}], the extended
envelope of  HD\,56126 shows no clear structure in the form of jets, cavities, etc.

The most recent result in the optical spectroscopy of PPNe is the discovery of the heavy-metal enrichment of the circumstellar
envelope of the V5112\,Sgr supergiant~[\cite{19500}]. The splitting of the profiles of the strongest heavy-metal
absorptions in the spectrum of this star with a complex bipolar envelope suggests that the process of the formation of a
structured circumstellar envelope may contribute to the enrichment of this envelope by products of stellar nucleosynthesis. The
profiles of the split lines contain a photospheric and two envelope components, one of which, like in the case of the 
CO--profile, arises in the envelope that formed at the AGB--stage  and expands at a velocity of V$_{\rm exp}(2)\approx 20$~km\,s$^{-1}$,
and the other one---in the envelope that formed later and expands at a velocity of V$_{\rm exp}(1)\approx 30$~km\,s$^{-1}$.  The
circumstellar components of strong heavy-metal absorptions have also been conclusively identified in the spectrum of
V354\,Lac~[\cite{V354Lac}]. Note that it is for these two related objects, V5112\,Sgr~\cite{19500} and
V354\,Lac~[\cite{V354Lac}], that we made the conclusion about the presence of circumstellar DIB-like absorptions in their
spectra.

The structure of the circumstellar nebula of  V354\,Lac may be more complex than it appears in the HST observations (see
Table\,\ref{PPN}). Polarimetric observations by Gledhill et al.~[\cite{Gledhill}] indicate the presence
of a ring structure embedded in an extended nebula. Nakashima et al.~[\cite{Nakashima2012}] point out that the axes of
the optical and infrared images of the nebula are almost perpendicular to each other. Based on the kinematic pattern of the
nebula as determined from the mapping of the CO--emission, Nakashima et al.~[\cite{Nakashima2012}] concluded that
the structure of the nebula includes not only a torus and a spherical component but also another element (possibly a jet).

The asymmetry (and, in particular, bipolarity) of the structure of selected PPNe  was found in several types of observations.
Fundamentally important information about the envelope morphology and dust grain properties in circumstellar envelopes can be
obtained from spectropolarimetric observations. It  follows from the observations of Bieging et al.~[\cite{Bieging}] that
more than 70\% of AGB and post--AGB stars have intrinsic polarization. The high linear polarization degree of their
radiation is in itself indicative of the asymmetry of the envelope. Examples include objects with bipolar envelopes, such as
RAFGL\,618~[\cite{Cohen}] and RAFGL\,2688~[\cite{Egg2}]. The linear polarization spectrum of RAFGL\,2688 taken with the 6--m 
telescope with a good spectral resolution (R\,=\,15\,000) in the 5000--6600~\AA{} wavelength interval  allowed the features 
of photospheric and circumstellar origin to be separated for the first time. Klochkova
et al.~[\cite{Egg2}], in particular, established that emission in the Swan system lines forms in the envelope and that
the corresponding transitions are excited by resonance fluorescence~[\cite{Egg2}].

Currently, no consensus has been reached concerning the development of deviations from spherical symmetry in PPNe. Ueta et
al.~[\cite{Ueta2000,Siodmiak}] analyzed high spatial resolution optical images of a sample of PPNe taken by the
Hubble Space Telescope and concluded that the optical depth of the circumstellar matter is the crucial factor that determines the
formation of a particular morphology of stellar envelopes. The dense and often spherical~[\cite{Castro2010}] envelope
that formed during the AGB--stage is believed to expand slowly, whereas the rapidly expanding feature is the axisymmetric part of
the envelope that formed later at the post--AGB stage. The sequence of these processes results in the development of an optical depth
gradient in the direction from the equator to the polar axis of the system. The presence of a companion and/or a magnetic field in
the system may also prove to be the physical factor that causes the loss of the spherical symmetry of the stellar envelope during
the short evolutionary interval between the AGB and post--AGB stages (see [\cite{Huggins,Ferreira}] and references therein). 
In their recent paper Koning et al.~[\cite{Koning}] proposed a simple PPN model based on a pair of evacuated cavities inside a 
dense spherical halo. The above authors demonstrated that all the morphological features observed in real bipolar PPNe can be 
reproduced by varying the available parameters (mass density inside the cavity, its size and orientation) of this model.

So far, the discovery of the heavy-metal enrichment of the circumstellar envelopes of the supergiants  V5112\,Sgr and V354\,Lac 
remains a unique result. It may be followed up by high-resolution spectroscopy of closely related objects. We believe 
the most promising objects to be IRAS\,04296+3429 and 23304+6147. It follows from Table\,\ref{PPN} that the atmospheres 
of the faint central stars of these sources are enriched in heavy metals, and their circumstellar envelopes have a complex structure.

\section{MAIN CONCLUSIONS}\label{conclus}

We used the results of high spectral resolution observations made with the 6--m telescope of the Special Astrophysical Observatory to
analyze the peculiarities of the optical spectra of a sample of post--AGB stars with atmospheres enriched in carbon and heavy
s--process metals and with carbon-enriched circumstellar envelopes.

We showed that the peculiarities of the line profiles  (the presence of an emission component in the lines of the Na\,I 
D--doublet, the type of the molecular features, the asymmetry and splitting of the profiles of strong absorptions with a low
excitation potential of the low level) are associated with the kinematic and chemical properties of the circumstellar envelope
and the type of its morphology. In particular, the variability of the observed profiles of  H$\alpha$ absorption and emission lines
and those of metal lines as well as the change of the type (absorption$/$emission) of  C$_2$ Swan bands observed in some
objects are associated with the changes in the structure of the circumstellar envelope.

The type of the H$\alpha$ profile (pure absorption, pure emission, P\,Cyg or inverse  P\,Cyg, with two emission components 
in the wings) is unrelated to the chemical composition of the atmosphere of the central star. The main factors that
influence the type of the H$\alpha$ profile and its variability are the mass loss rate, stellar wind velocity, kinematics and 
optical depth of the envelope.

The splitting of the profiles of the strongest heavy-metal absorptions in the spectra of the  V5112\,Sgr  and V354\,Lac supergiants 
found as a result of our observations suggests that the formation of a structured circumstellar envelope is accompanied by the 
enrichment of this envelope with the products of stellar nucleosynthesis. Note that it is in the spectra of these two related objects,
V5112\,Sgr  and V354\,Lac, that we found DIB--like envelope absorptions.

Attempts to find a relation between the peculiarities of the optical spectrum and the morphology of the circumstellar medium 
are complicated by the fact that the observed structure of the envelope depends strongly on the inclination of the symmetry axis
to the line of sight and on the angular resolution of the spectroscopic and direct-imaging instruments.

\section*{Acknowledgments}

I am deeply grateful to all the coauthors who took part in carrying out the program of spectroscopic observations of
supergiants with IR excesses at the SAO 6--m telescope. This work was supported by the Russian Foundation for Basic Research
(project No.\,Envelopes14--02--00291\,a). This research has made use of the SIMBAD database, operated at CDS, Strasbourg, France, and NASA
Astrophysics Data System. Observations at the 6--m telescope of the Special Astrophysical Observatory are carried out with the
financial support of the Ministry of Education and Science of the Russian Federation (contracts No.~14.518.11.7070 and
16.518.11.7073).

\end{document}